\documentclass[aip,reprint]{revtex4-1}
\usepackage{graphicx}
\begin{document}

\title{Metasurface polarization splitter}

\author{Brian A. Slovick}\email{Corresponding author: brian.slovick@sri.com}
\affiliation{Applied Optics Laboratory, SRI International, Menlo Park, California \mbox{94025, United States}}
\author{You Zhou}
\affiliation{Interdisciplinary Materials Science Program, Vanderbilt University, \mbox{Nashville, Tennessee 37212, United States}}
\author{Zhi Gang Yu}
\affiliation{Institute for Shock Physics, Washington State University, Pullman, Washington \mbox{99164, United States}}
\author{Ivan I. Kravchenckou}
\author{Dayrl P. Briggs}
\affiliation{Center for Nanophase Materials Sciences, Oak Ridge National Laboratory, Oak Ridge, Tennessee 37831, United States}
\author{Parikshit Moitra}
\affiliation{Optoelectronics Research Centre, University of Southampton, Southampton \mbox{SO17 1BJ, United Kingdom}}
\author{Srini Krishnamurthy}
\affiliation{Applied Optics Laboratory, SRI International, Menlo Park, California \mbox{94025, United States}}
\author{Jason Valentine}
\affiliation{Department of Mechanical Engineering, Vanderbilt University, Nashville, Tennessee \mbox{37212, United States}}

\begin{abstract}
Polarization beam splitters, devices that separate the two orthogonal polarizations of light into different propagation directions, are one of the most ubiquitous optical elements. However, traditionally polarization splitters rely on bulky optical materials, while emerging optoelectronic and photonic circuits require compact, chip-scale polarization splitters. Here we show that a subwavelength rectangular lattice of cylindrical silicon Mie resonators functions as a polarization splitter, efficiently reflecting one polarization while transmitting the other. We show that the polarization splitting arises from the anisotropic permittivity and permeability of the metasurface due to the two-fold rotational symmetry of the rectangular unit cell. The high polarization efficiency, low loss, and low profile make these metasurface polarization splitters ideally suited for monolithic integration with optoelectronic and photonic circuits.
\end{abstract}
\maketitle

In general, polarizers can be divided into two types: absorptive polarizers and polarization beam splitters. Absorptive polarizers use wire grids, dichroic materials, or nanoparticle composites to absorb the rejected polarization.\cite{Saleh2007,Nicolais2004} Although they provide high degrees of polarization, because the rejected polarization is absorbed, both polarizations cannot be analyzed simultaneously. This makes them unsuitable for applications such as optical quantum computing, where polarization splitters are needed to produce quantum bits, or qubits, by dividing single circularly-polarized photons into a superposition of vertical and horizontal polarizations.\cite{Pittman2001,Pan2003,Politi2008,Gevaux2008} Alternatively, polarization beam splitters preserve the rejected polarization either by reflection or diffraction. Commercially available polarization beam splitters separate the polarizations using total internal reflection in birefrigent cubes or Brewster angle reflection in multilayer dielectric films. While these devices provide efficient polarization splitting, their large profile makes them incompatible with chip-scale photonic and optoelectronic devices.\cite{Walker1993,Wang2007}

Recently, there has been considerable effort to redesign bulky optical elements using metasurfaces,\cite{Kildishev2013,Yu2014} i.e., planar, subwavelength layers with structural elements designed to modify the amplitude, phase, and polarization of scattered light. While originally developed with metallic resonators,\cite{Yu2011,Ni2012,Sun2012,Huang2012,Pors2013,Aieta2012} recent studies have shown that dielectric resonators can be used to realize lower absorption loss. Demonstrations of all-dielectric metasurfaces include metasurface reflectors\cite{Moitra2014,Moitra2015,Liu2014} and antireflection coatings,\cite{Spinelli2012} Fano-resonant surfaces for narrowband spectral filtering,\cite{Miroshnichenko2012,Wu2014,Zhang2013} gradient-phase metasurfaces functioning as spatial-light modulators \cite{Yang2014,Chong2015,Lin2014} and lenses,\cite{Aieta2015}  and periodic arrays of anisotropic (i.e., polarization-dependent) scattering elements functioning as polarizers,\cite{Shen2014} polarization rotators,\cite{Yang2014} and converters.\cite{Wu2014}

Gradient-phase metasurfaces have also been used to design polarization splitters.\cite{Walker1993,Wang2007,Zheng2014,Arbabi2015} These function by diffracting the two orthogonal polarizations into different grating modes using periodic binary or multilevel phase gratings with anisotropic unit cells. However, their performance is limited by diffraction efficiency, as some energy is always lost to the specular mode due to diffraction associated with the finite length of the unit cell.\cite{Walker1993,Suleski2004} Alternatively, integrated polarization splitters have been designed for waveguide propagation using genetic algorithms,\cite{Shen2015} but their efficiencies are low, and the tedious design process is not readily adaptable to different wavelengths or materials.

In contrast to these designs, our polarization splitter derives its properties from an anisotropic permittivity and permeability originating from a subwavelength rectangular lattice. We showed previously that a subwavelength \textit{square} array of silicon cylinders can reflect more than 99.7\% of unpolarized near infrared light.\cite{Slovick2013,Moitra2014,Moitra2015} Such high reflectivity by a subwavelength layer is achieved by choosing the cylinder size to obtain enhanced backscattering due to electric and magnetic Mie resonances. By breaking the four-fold rotational symmetry of the square lattice, here we show that a \textit{rectangular} lattice with two-fold rotational symmetry can be used to make near-infrared beam-splitting polarizers, efficiently reflecting one polarization while transmitting the other. Owing to the subwavelength lattice, our polarization splitter behaves like a homogeneous medium with only one propagating mode per polarization, and thus is not limited by diffraction efficiency. Along with particle size, the additional degree of freedom provided by the anisotropic lattice allows us to independently tailor the reflection for the two polarizations, and paves the way for a new class of metasurface polarization elements.

\begin{figure}
\centering
\includegraphics[width=62mm]{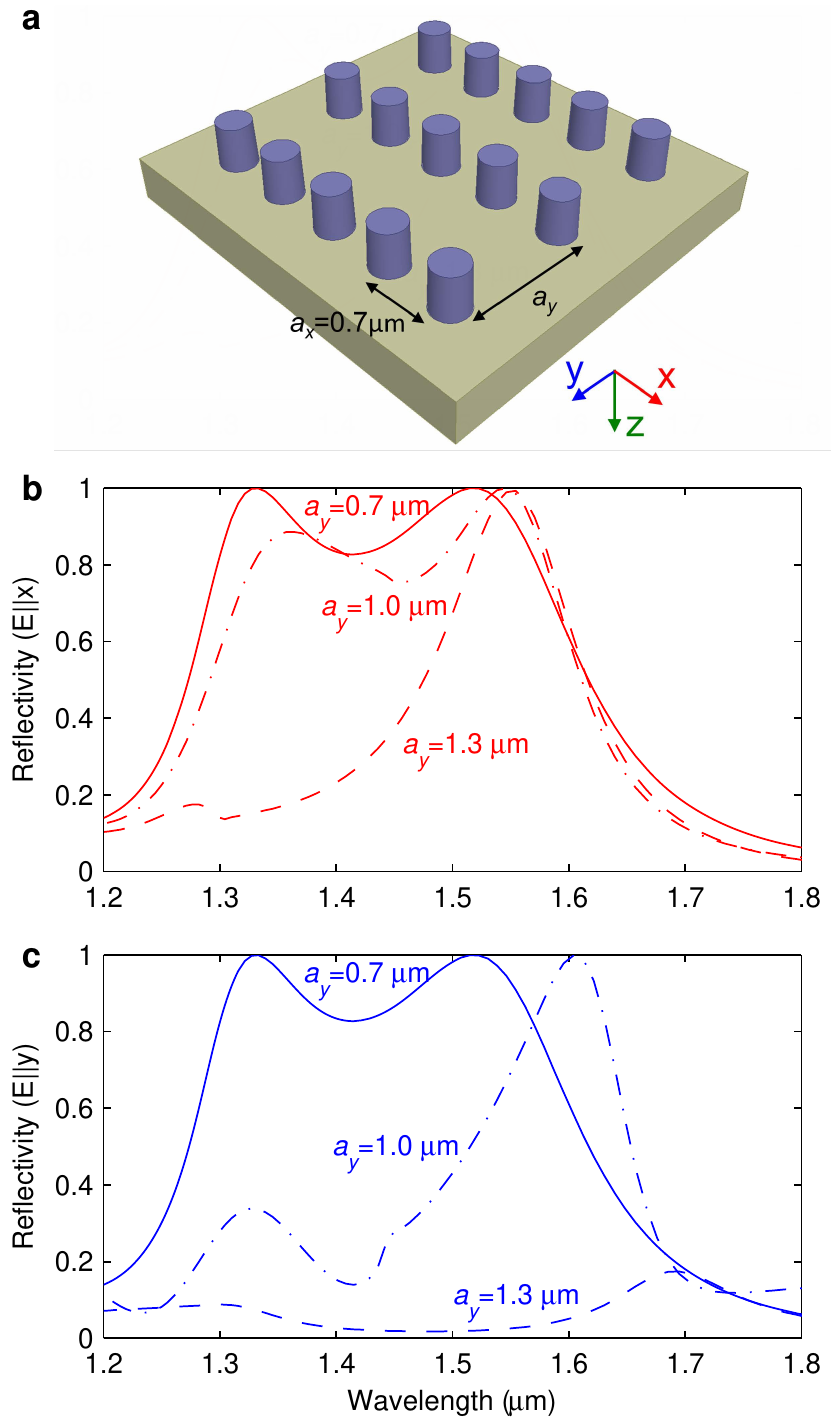}
\caption{Schematic of the anisotropic metasurface (\textbf{a}) and the calculated reflectivity for $a_x=0.7$ $\mu$m and different values of $a_y$, for light propagating along z and polarized along x (\textbf{b}) and y (\textbf{c}). As $a_y$ increases, the reflection maximum due to the electric resonance at 1.3 $\mu$m decreases for both polarizations, while the maximum due to the magnetic resonance at 1.55 $\mu$m decreases only for $\mathbf{E}||\mathbf{y}$.}
\label{fig1}
\end{figure}

First, we apply our full-wave models to show that a metasurface with a rectangular lattice functions as a polarization splitter (see Section 1 in Supplement 1). A schematic of the metasurface and the calculated polarization-dependent reflectivity for several rectangular lattices are shown in Figure \ref{fig1}. The metasurface consists of a rectangular array of silicon cylinders on silica substrate. The diameter and height of the cylinders, respectively, are 0.36 $\mu$m and 0.46 $\mu$m. To study the effects of a rectangular lattice, we fixed the periodicity along x to $a_x=0.7$ $\mu$m and increased the periodicity along y ($a_y$). For the isotropic square lattice with $a_y=a_x$, the reflectivity is equivalent for the two polarizations, with local maxima at 1.55 and 1.3 $\mu$m, corresponding to the magnetic and electric Mie resonances, respectively.\cite{Moitra2014,Moitra2015} As $a_y$ increases, the reflectivity at the electric resonance decreases for both polarizations, while the reflectivity at the magnetic resonance decreases only when the electric field is along the long axis of the rectangular unit cell ($\mathbf{E}||\mathbf{y}$). Thus, near the magnetic resonance the metasurface functions as a polarization splitter, efficiently reflecting one polarization while transmitting the other.

\begin{figure}[t]
\centering
\includegraphics[width=59.6mm]{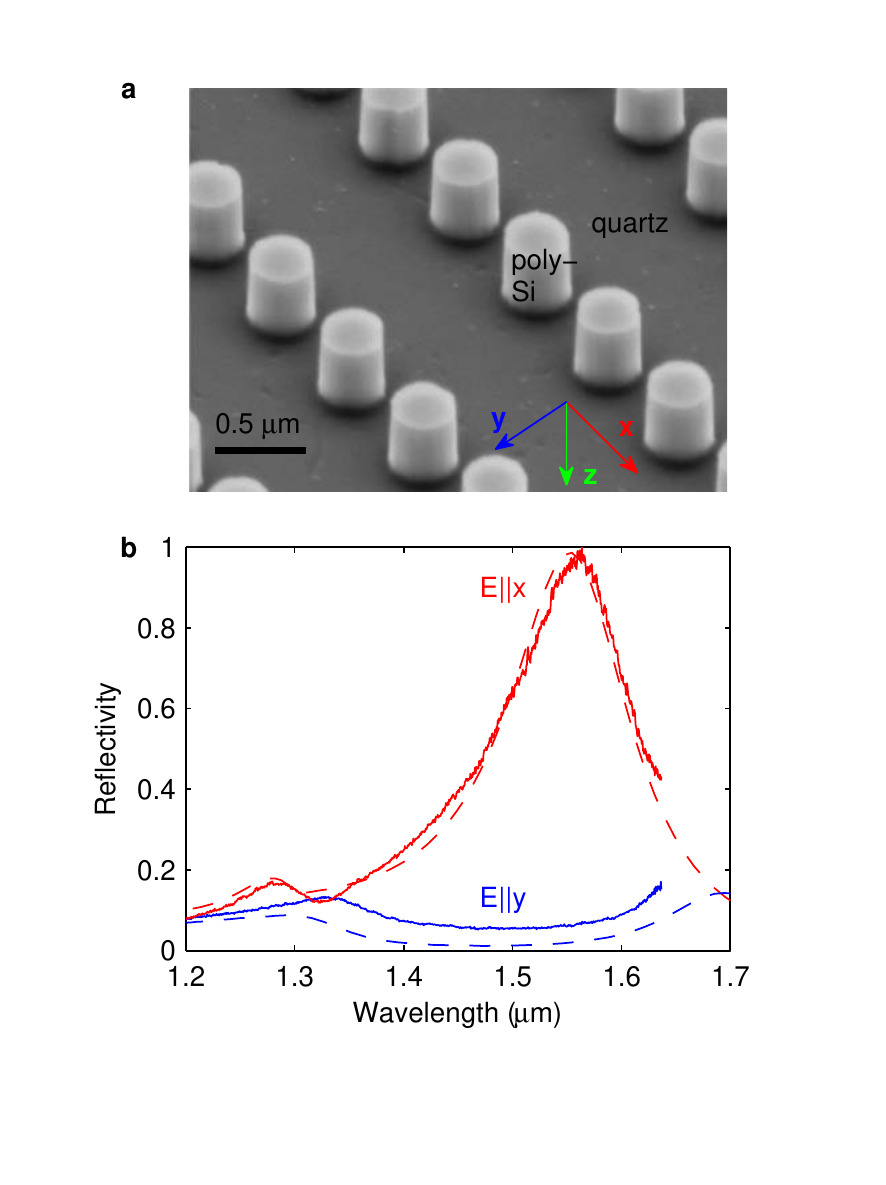}
\caption{SEM of the metasurface consisting of a rectangular array of poly-Si cylinders on quartz substrate (\textbf{a}) and the measured (solid lines) and modeled (dashed lines) reflectivity for light propagating along z and polarized along x and y (\textbf{b}). At 1.55 $\mu$m, the anisotropic metasurface efficiently reflects light polarized along x while transmitting light polarized along y, and thus functions as a polarization splitter.}
\label{fig2}
\end{figure}

The optimal design with $a_y=1.3$ $\mu$m was selected for fabrication. Figure \ref{fig2} shows a scanning electron micrograph (SEM) of the fabricated metasurface along with the measured and modeled reflectivity (see Section 3 in Supplement 1). The metasurface was fabricated from polycrystalline silicon (poly-Si) on silica substrate using electron beam lithography and reactive ion etching (see Section 2 in Supplement 1). From the SEM, the periodicity along x and y, respectively, is 0.7 and 1.3 $\mu$m, and the height, top diameter, and bottom diameter are 0.46, 0.35, and 0.37 $\mu$m. At the design wavelength of 1.55 $\mu$m, the predicted reflectivity is greater than 99\% for $\mathbf{E}||\mathbf{x}$, and less than 2\% for $\mathbf{E}||\mathbf{y}$. While the calculated reflectivity for $\mathbf{E}||\mathbf{x}$ is in excellent agreement with the measured value, the apparent discrepancy for $\mathbf{E}||\mathbf{y}$ can be explained by a slight (45 nm) overetch of the cylinders (see Section 4 in Supplement 1). The calculated transmission at 1.55 $\mu$m for $\mathbf{E}||\mathbf{y}$, shown in Section 4 of Supplement 1, is greater than 98\%. However, because the periodicity along y is larger than the design wavelength in the quartz substrate (1.07 $\mu$m), the substrate supports higher-order diffraction modes, namely $\pm1$ orders. We find that a significant fraction (40\%) of the transmitted power is diffracted into these higher-order modes. However, our models indicate that when the quartz substrate is replaced by porous Si, which does not support higher-order modes due to its low refractive index of 1.1, the zero-order transmission can be greater than 90\%.

\begin{figure}
\includegraphics[width=68mm]{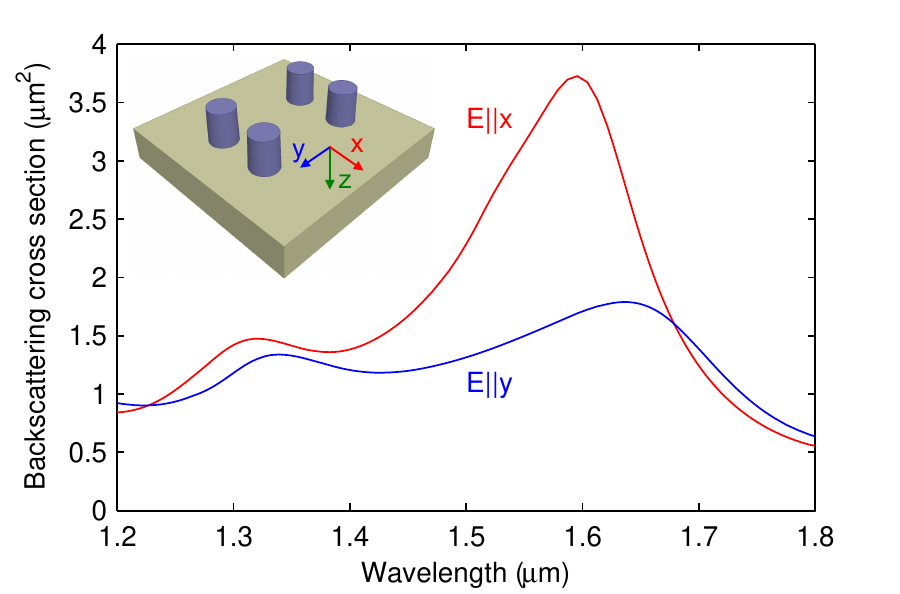}
\caption{Calculated backscattering cross section of a four-particle unit cell, showing that the reflectivity of the metasurface originates from the polarization-dependent resonant backscattering of the unit cell.}
\label{fig3}
\end{figure}

In contrast to interparticle Bragg scattering in diffraction gratings and photonic crystals, the high reflectivity of our metasurface originates from scattering resonances within the unit cell. This can be seen by analyzing the backscattering cross section of the four-particle unit cell, shown in Figure \ref{fig3}. We find that the maxima in the backscattering cross section of the unit cell coincide with the maxima in the reflectivity of the array. Noting that the scattering resonances near 1.55 $\mu$m and 1.3 $\mu$m, respectively, correspond to magnetic and electric modes, we find that only the magnetic mode for $\mathbf{E}||\mathbf{x}$ leads to a cross section considerably larger than the area of the unit cell (0.91 $\mu$m$^2$).

\begin{figure}[t]
\includegraphics[width=88mm]{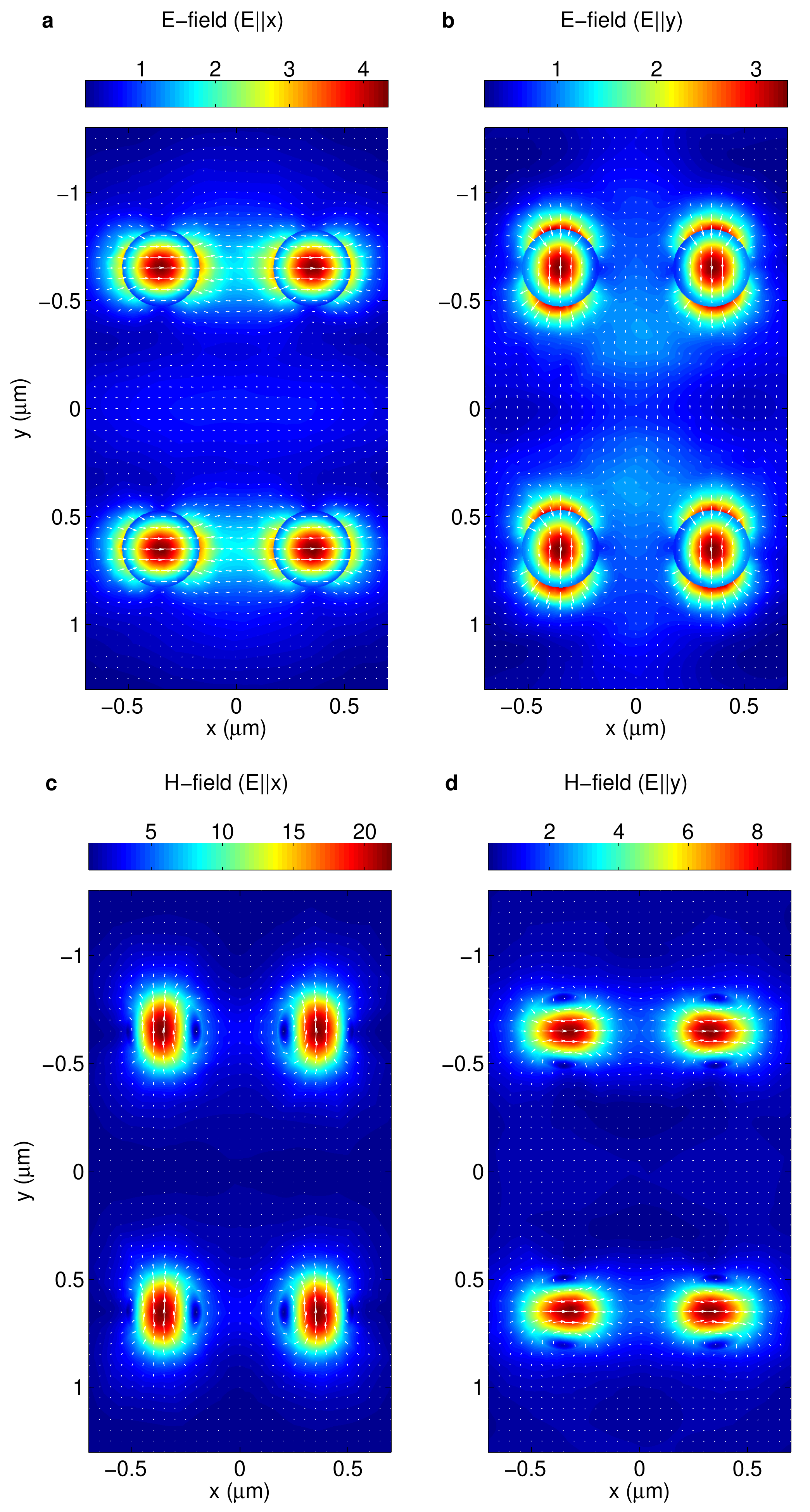}
\caption{Electric field enhancement at resonance (1.3 $\mu$m) for $\mathbf{E}||\mathbf{x}$ (\textbf{a}) and $\mathbf{E}||\mathbf{y}$ (\textbf{b}). Magnetic field enhancement at resonance (1.55 $\mu$m) for $\mathbf{E}||\mathbf{x}$ (\textbf{c}) and $\mathbf{E}||\mathbf{y}$ (\textbf{d}). The electric field is poorly confined for both polarizations, leading to relatively small field enhancement. The magnetic fields are well confined, particularly for $\mathbf{E}||\mathbf{x}$.}
\label{fig4}
\end{figure}

Since the cross section is proportional to the scattered field intensity, to understand why only the magnetic mode for $\mathbf{E}||\mathbf{x}$ leads to a large cross section and high reflectivity, we calculated the scattered electric and magnetic field distributions (normalized to the incident fields) at the respective Mie resonance frequencies for both polarizations (Figure \ref{fig4}). We find considerable electric field coupling between the cylinders for both polarizations, leading to relatively small field enhancement. On the other hand, the magnetic fields are well confined, particularly for $\mathbf{E}||\mathbf{x}$. The reason for the varying confinement of the electric and magnetic fields can be traced to their different boundary conditions. Since the normal components of \textbf{B} (=$\mu$\textbf{H}) and \textbf{D} (=$\epsilon$\textbf{E}) must be continuous across the cylinder surface, the normal component of \textbf{H} is also continuous because $\mu=1$ throughout, whereas the normal component of \textbf{E} is discontinuous owing to the large dielectric mismatch between silicon and free space. This leads to a larger normal component of $\mathbf{E}$ just outside the cylinder surface, and hence electric field coupling and poor confinement. Alternatively, the normal component of $\mathbf{H}$ is continuous, leading to better field confinement and less coupling, particularly when the magnetic field is perpendicular to the short axis of the rectangular unit cell (i.e., $\mathbf{E}||\mathbf{x}$). The large magnetic field enhancement and weak interparticle coupling for $\mathbf{E}||\mathbf{x}$ leads to a large backscattering cross section and reflectivity at 1.55 $\mu$m.

\begin{figure}[t]
\includegraphics[width=78mm]{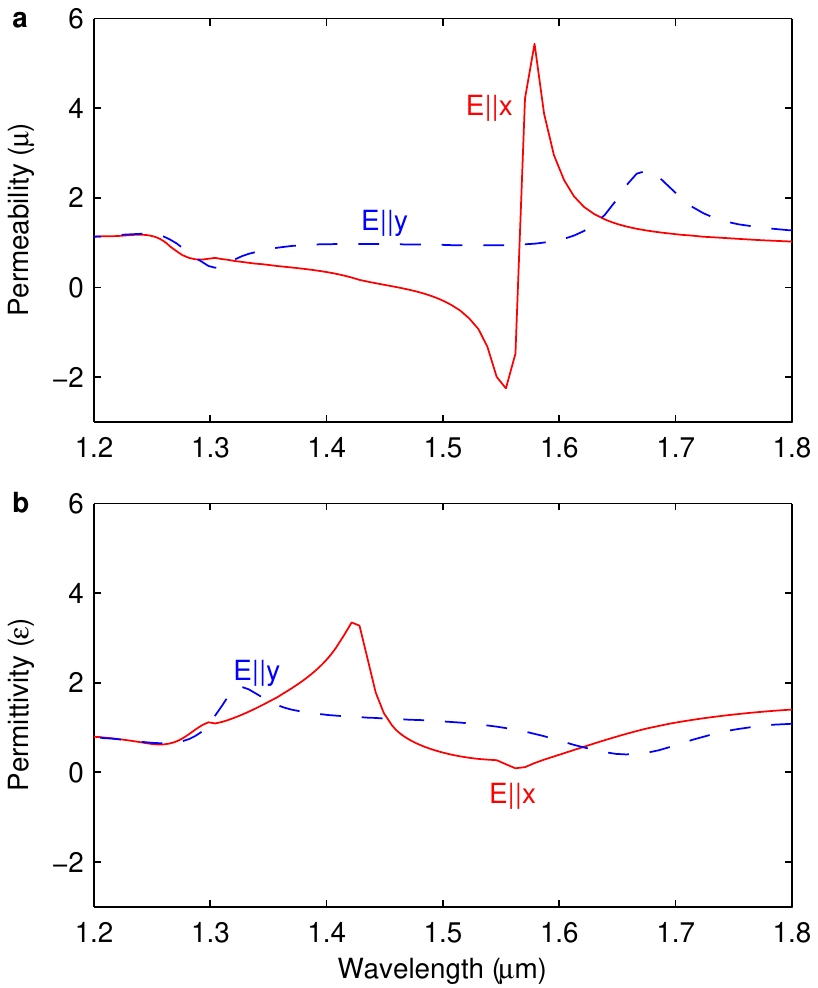}
\caption{Effective permeability (\textbf{a}) and permittivity (\textbf{b}) of the anisotropic metasurface polarization splitter. At 1.55 $\mu$m the magnetic resonance for $\mathbf{E}||\mathbf{x}$ leads to negative $\mu$, while both $\epsilon$ and $\mu$ for $\mathbf{E}||\mathbf{y}$ are close to 1.}
\label{fig5}
\end{figure}

Owing to the subwavelength lattice spacing (with respect to free space), our polarization splitter supports only one propagating mode per polarization, and thus behaves like a homogeneous medium. Therefore, the polarization-dependent reflection can be understood in terms of the anisotropic effective parameters of the metasurface. Figures \ref{fig5}a and b, respectively, show the effective permeability ($\mu$) and permittivity ($\epsilon$) of the anisotropic metasurface calculated using $S$-parameter inversion (see Section 1 in Supplement 1).\cite{Smith2002} We showed previously that bands of high reflectivity occur where $\epsilon$ or $\mu$ is negative,\cite{Slovick2013} in contrast to negative index materials, which require both $\epsilon$ and $\mu$ to be negative. We find that only the magnetic resonance for $\mathbf{E}||\mathbf{x}$ leads to a negative $\mu$, in this case around 1.55 $\mu$m, coinciding with the peak reflectivity in Figure \ref{fig2}b. At the same wavelength, we find that both $\epsilon$ and $\mu$ for $\mathbf{E}||\mathbf{y}$ are close to 1, consistent with the low reflectivity of the metasurface at 1.55 $\mu$m for $\mathbf{E}||\mathbf{y}$.

In summary, we have shown that a subwavelength rectangular lattice of cylindrical silicon Mie resonators functions as a polarization splitter, efficiently reflecting one polarization while transmitting the other. The polarization-dependent reflection arises from anisotropic near-field coupling between resonators in the rectangular lattice, and can be understood in terms of the anisotropic permittivity and permeability of the metasurface. The polarization efficiency is considerably larger than for devices based on diffraction \cite{Zheng2014,Arbabi2015} and comparable to commercial polarizing beamsplitter cubes. The high degree of polarization, low loss, and low profile of these metasurface polarization splitters may lead to novel designs for integrated photonics and chip-scale optical quantum devices. Most importantly, the introduction of an additional degree of freedom--the anisotropic lattice--paves the way for a new class of metasurface polarization elements.
\\

\textbf{\uppercase{Funding Information}}

Office of Naval Research (ONR) (N00014-12-1-0722, N00014-14-1-0475).
\\

\textbf{\uppercase{Acknowledgment}}

A portion of this research was conducted at the Center for Nanophase Materials Sciences, which is a DOE Office of Science User Facility.
\\

\textbf{\uppercase{Supplemental Documents}}

See Supplement 1 for supporting content.

\bibliography{bib}

\end{document}